\documentstyle[aps,prl,twocolumn]{revtex}

\begin{document}
\title{Incipient superconductivity in TaB$_{2}$}
\author{D. Kaczorowski, A. J. Zaleski, O. J. \.{Z}oga\l , and J. Klamut \cite{ML}}
\address{Institute of Low Temperature and Structure Research, Polish Academy of
Sciences,\\ P.O.Box 1410, 50-950 Wroc\l aw, Poland}
\date{\today}

\maketitle

\begin{abstract}
Magnetic properties of TaB$_{2}$ were studied by means of DC and
AC susceptibility measurements. It has been found that the
compound becomes superconducting at T$_{c}$ = 9.5 K. From upper
critical field measurements the superconducting parameters have
been evaluated, which indicated that TaB$_{2}$ is a hard type-II
superconductor. Possible reasons why superconductivity has not
been discovered in previous studies are briefly discussed.
\end{abstract}


\bigskip

\section{Introduction}

The recent discovery of high-T$_{c}$ superconductivity in
MgB$_{2}$ \cite {Nagamatsu} has raised considerable interest in a
search for analogous behavior in similar systems. Although the
most natural candidates for such investigations seem to be
AlB$_{2}$-type alkali, alkali earth or group III - element
diborides, AB$_{2}$ (e.g. A = Li, Be, Al), in none of them
superconductivity has been found up to now
\cite{Slusky}\cite{Felner}\cite {Zhao}. Moreover, in the solid
solutions Mg$_{1-x}$A$_{x}$B$_{2}$ a rapid decrease of
superconducting transition temperature is always observed with the
increase in the dopant content \cite{Slusky}\cite{Zhao}.

Diborides isostructural to MgB$_{2}$ form also with transition
metals from the IVa, Va and VIa groups of the Periodic Table (T =
Ti, Zr, Hf, V, Nb, Ta, Cr, Mo), yet also these compounds have been
reported \cite{Leyarovska} not to show superconductivity down to
0.42 K, with the only exception for NbB$_{2}$ that becomes
superconducting below T$_{c}$ = 0.62 K.

Amidst the TB$_{2}$ materials, TaB$_{2}$ seems to be especially
interesting due to a unique feature of its (0001) surface being
terminated by a graphitic boron layer \cite{Kawanowa}. The recent
electronic structure calculations of the TaB$_{2}$ (0001) surface
have revealed the presence at the Fermi level of a pronounced DOS
peak originated predominantly from the B {\it 2p} orbitals
\cite{Kawanowa2}. Due to a strong charge transfer from Ta to B the
Ta {\it 5d} orbitals are lowered in energy and filled with
electrons. Interestingly, the resulting band structure is similar
to the one derived for MgB$_{2}$ \cite{Satta}\cite{Medvedeva} and
favorable of hole superconductivity concept \cite{Hirsch}. This
striking observation motivated us to re-investigate our TaB$_{2}$
sample on the context of possible appearance of superconductivity.

\section{Experimental}

Powder sample of TaB$_{2}$ was obtained by the borothermic method
\cite {Peshev}. The product was checked by chemical analysis,
which proved a sample stoichiometry close to the ideal. EDAX
studies, performed using a Phillips 515 scanning electron
microscope did not reveal any other elements but tantalum. Phase
analysis showed a homogeneous material. X-ray diffraction pattern
(see Fig. 1) was easily indexed within a hexagonal unit cell with
lattice parameters: {\it a} = 308.2 pm and {\it c} = 324.3 pm. DC
magnetic measurements were carried out using a Quantum Design
SQUID magnetometer. The AC susceptibility was measured employing
an Oxford Instruments EXA susceptometer. All physical measurements
were done on powders freely placed in sample holders.

\section{Results}

In Fig. 2 is shown the temperature dependence of the magnetization in TaB$%
_{2}$ measured in a field of 50 Oe upon cooling the sample in zero
(ZFC) and applied (FC) magnetic field. Down to T$_{c}$ = 9.5 K,
the specimen investigated shows nearly temperature independent
paramagnetism with the magnetic susceptibility of the order of
10$^{-5}$ emu/g. However, most strikingly, at the temperature
T$_{c}$ there occurs a clear transition to a diamagnetic state,
and the ZFC and FC curves split in a manner characteristic of
type-II superconductors, with the weak irreversibility field.
Furthermore, the field variation of the magnetization, taken at
1.7 K with increasing and decreasing magnetic field (see Fig. 3),
shows highly irreversible properties of the material. The lower
critical field estimated from Fig. 3 is about H$_{c1}$= 100 Oe.

The behavior typical of type-II superconductors has been
corroborated for TaB$_{2}$ via the AC magnetic susceptibility
measurements (H = 10 Oe, f = 1 KHz). As seen in Fig. 4, in the
normal state the real component of the susceptibility is positive
and below about 10 K a sudden drop to negative values occurs. Just
below T$_{c}$ a sharp peak in the imaginary component of the
susceptibility is observed, being characteristic of good quality
superconducting material.

The onset of superconductivity in the AC susceptibility was used
to define the upper critical field H$_{c2}$, and the
so-derived temperature dependence of H$_{c2}$ is depicted in Fig. 5. The experimental data can be well described by the expression: H$_{c2}$(T) = H$_{c2}$%
(0)[1-T/T$_{c}$]$^{\beta }$, with the least-squares fitting
parameters: H$_{c2}$(0) = 23.6 kOe and $\beta $ = 2.2. It is
worthwhile noting that the H$_{c2}$(T) variation has a positive
curvature, yielding a rather large parameter $\beta $, which is
usually considered as a measure of the material quality. The
H$_{c2}$(T) dependence derived for TaB$_{2}$ differs from that
known for conventional low-T$_{c}$ superconductors, but rather
resembles the behavior typical of rare-earth nickel borocarbides
\cite{Fuchs}.

>From the obtained value of H$_{c2}$(0) (which is probably the
upper estimate of the real value) it is possible to derive the
superconducting coherence length $\xi _{0}$ = 13.4 nm. Then, from
H$_{c1}$ and $\xi _{0}$ we estimate the penetration depth $\lambda
_{0}$ = 250 nm and the Ginzburg-Landau parameter $\kappa $ = 18.
All these superconducting characteristics strongly indicate that
TaB$_{2}$ may be classified as a hard type-II superconductor.

\section{Discussion}

The discovery of superconductivity in TaB$_{2}$ below T$_{c}$ =
9.5 K clearly contradicts previous results obtained for this
compound by Leyarovska and Layarovski \cite{Leyarovska}. It should
be emphasized that we have also re-investigated some other
transition metal diborides studied by the latter authors
(TiB$_{2}$, ZrB$_{2}$, HfB$_{2}$, VB$_{2}$, NbB$_{2}$) but only
confirmed their conclusions that none of them is superconducting
above 1.7 K. Thus, assuming that the previously investigated
non-superconducting sample of TaB$_{2}$ was of the highest
quality, we thoroughly checked our specimen on possible
impurities, which might result in spurious superconductivity
effect.

As can be inferred from Fig. 1, two very weak impurity lines were
indeed observed in the X-ray pattern. Both can be ascribed to the
equiatomic compound TaB or alternatively, but less likely, to
Ta$_{3}$B$_{4}$. In order to test a hypothesis that the
superconductivity arises due to the presence of small amount of
the other tantalum borides, we prepared both these phases and
investigated their low-temperature magnetic and electrical
properties. Although one of them is diamagnetic, none has proven
to be superconducting.

Then, we considered a possibility that the amount of eventual
superconducting impurity in our sample is below the detection
limit of X-ray diffraction. This latter presumption seems
supported by rather weak diamagnetic signal found in the
superconducting state. The contamination by non-reacted metallic
Ta must be ruled out because the afore-discussed critical
temperature and upper critical field are very different than those
characterizing pure Ta (T$_{c}$ = 4.4 K and H$_{c2}$ = 0.8 kOe).
Although the detailed EDAX measurements of our sample has not
revealed the presence of any other elements except Ta, we cannot
exclude that it does contain some oxygen or other light elements,
which are not seen by EDAX. In such a case the observed
superconductivity could be due to very little amount of unknown
tantalum oxide or e.g. nitride. Yet, no other Ta-based compounds,
reported so far in the literature, fit to the behavior found in
the present work. In particular, albeit the superconducting
temperature for the tantalum monocarbide TaC is quite similar to
T$_{c}$ found for our sample (9-11.4 K \cite{Wells}), the reported
H$_{c2}$ is only 4.6 kOe \cite{Fink}.

Alternatively, the apparent discrepancy in the properties of our
specimen of TaB$_{2}$ and that studied in Ref. \cite{Leyarovska}
could be attributed to some difference in the stoichiometry of
these two samples. The hexagonal phase TaB$_{2}$ is known to form
within a rather broad homogeneity range \cite{Massalski} giving
sub- and superstoichiometric compositions. It seems conceivable
that the effect of superconductivity is restricted to a particular
nonstochiometric composition TaB$_{2\pm x}$ and is governed by existing defects in the AlB$%
_{2}$-type unit cell, which can play similar role as oxygen in
high-T$_{c}$ superconductors.

\acknowledgments The authors are indebted to Dr. P. Peshev for
providing the sample of TaB$_{2}$ and Dr. Z. Bukowski for
synthesizing the samples of TaB and Ta$_{3}$B$_{4}$.

\subsection{Figure captions}

Fig. 1. Powder X-ray diffraction pattern (CuK$\alpha _{1}$ radiation) for TaB%
$_{2}$ at room temperature. Arrows mark the reflections from
possible TaB impurity.

Fig. 2. Low-temperature dependence of the magnetization in TaB$_{2}$, taken
at H = 50 Oe in the ZFC and FC regimes.

Fig. 3. Field variation of the magnetization in TaB$_{2}$,
measured at T = 1.7 K with increasing and decreasing field.

Fig. 4. Temperature dependence of the in-phase and out-of-phase
magnetic susceptibility components for TaB$_{2}$, measured in the
AC field H = 10 Oe with the frequency f = 1 kHz.

Fig. 5. Temperature variation of the upper critical field H$_{c2}$
in TaB$_{2}$. The solid line is a fit of the experimental data to
the function H$_{c2}$ = H$_{c2}$%
(0)[1-T/T$_{c}$]$^{\beta }$ with the parameters: H$_{c2}$(0) =
23.6 kOe and $\beta $ = 2.2.

\end{document}